# Quantum metric of non-Hermitian Su-Schrieffer-Heeger systems

Chao Chen Ye,[1,2,*] W. L. Vleeshouwers,[1,3,†] S. Heatley,[3,‡] V. Gritsev,[3,§] and C. Morais Smith[1,‖]
[1]*Institute for Theoretical Physics, Utrecht University, Princetonplein 5, Utrecht, 3584 CC, Netherlands*
[2]*Zernike Institute for Advanced Materials, University of Groningen, Nijenborgh 4, 9747 AG Groningen, The Netherlands*
[3]*Institute for Theoretical Physics, University of Amsterdam, Science Park 904, Postbus 94485, 1098 XH Amsterdam, The Netherlands*

Topological insulators have been studied intensively over the last decades. Earlier research focused on Hermitian Hamiltonians, but recently, peculiar and interesting properties were found by introducing non-Hermiticity. In this work, we apply a quantum geometric approach to various Hermitian and non-Hermitian versions of the Su-Schrieffer-Heeger (SSH) model. We find that this method allows one to correctly identify different topological phases and topological phase transitions for all SSH models, but only when using the metric tensor containing both left and right eigenvectors. Whereas the quantum geometry of Hermitian systems is Riemannian, introducing non-Hermiticity leads to pseudo-Riemannian and complex geometries, thus significantly generalizing from the quantum geometries studied thus far. One remarkable example of this is the mathematical agreement between topological phase transition curves and lightlike paths in general relativity, suggesting a possibility of simulating space-time in non-Hermitian systems. We find that the metric in non-Hermitian phases degenerates in such a way that it effectively reduces the dimensionality of the quantum geometry by one. This implies that within linear response theory, one can perturb the system by a particular change of parameters while maintaining a zero excitation rate.

## I. INTRODUCTION

The discovery of topological insulators is one of the most important developments in condensed-matter physics of the past decades [1–4]. Their peculiar properties, in particular the coexistence of an insulating bulk with quantized gapless edge states that are robust against perturbations, hold promises to revolutionize our current technology. An elegant classification of the topological phases based on symmetries and dimensionality has been proposed for noninteracting systems [5], leading to the tenfold way. Later, this classification was extended to out-of-equilibrium and current efforts focus on interactions. The mechanisms at work in topological systems have since then been generalized to mechanical systems [6] and even to the Earth itself [7].

A partly contemporary development is the growing importance of non-Hermitian (NH) quantum systems, both in theory and experiment. Non-Hermiticity can arise as an effective description of open quantum systems, e.g., driven and/or dissipative ones. When quantum jumps can be neglected, they can be described by NH Hamiltonians. In general, NH Hamiltonians can have complex energy eigenvalues that may appear in various forms, depending on the system's symmetries. In particular, the eigenvalues of $\mathcal{PT}$-symmetric NH Hamiltonians are real when $\mathcal{PT}$-symmetry is preserved in the topological phase, but come in complex-conjugate pairs when this symmetry is spontaneously broken [8]. The eigenvalues of pseudo-Hermitian Hamiltonians $\hat{H}$, for which there exists an invertible operator $\eta$ such that $\hat{H} = \eta^{-1}\hat{H}^\dagger \eta$, appear in complex-conjugated pairs. An essential feature of NH operators is the fact that their left and right eigenvectors are no longer related by Hermitian conjugation. Perhaps the most striking feature of NH systems is the strong sensitivity of the bulk to boundary conditions. This is known as the NH *skin effect*, which has been explained in terms of an intrinsic NH topology [9]. For a NH lattice model with nonreciprocal real-valued hoppings, the skin effect can simply be understood as the accumulation of states at a boundary of the system. The generalization of the tenfold way to NH topological insulators exhibits an intricate classification structure [10,11], and the additional NH parameters lead to enriched phase diagrams as compared to the Hermitian case [12].

Central to the understanding of Hermitian topological insulators is the antisymmetric part of the quantum geometric tensor (QGT), that is, the Berry curvature. From the Berry curvature, one can calculate the first Chern number of the $U(1)$-bundle, which counts the number of gapless edge modes. The symmetric part of the QGT, the quantum metric tensor (QMT), was introduced by Provost and Vallee [13]. It provides a natural measure for the distance between quantum states, and as such, plays an important role in quantum information theory [14]. In photonics, the QMT becomes a crucial quantity when determining wave packets dynamics near exceptional points for NH systems [15]. For flat-band systems, the superfluidity and stable supercurrents are possible if the band has a nontrivial quantum geometry: superfluid weight[1] is connected to the quantum metric and bounded below by a finite Chern number, while superconductivity is directly governed by the QMT [14].

[*]c.chen.ye@rug.nl
[†]w.l.vleeshouwers@uva.nl
[‡]sarahmheatley@gmail.com
[§]v.gritsev@uva.nl
[‖]C.deMoraisSmith@uu.nl

The QMT in essence describes a pull-back of the Fubini-Study metric of the Hilbert space to a manifold of parameter space. It defines a geodesic distance between points on the parameter manifold and serves as a measure of the dissimilarity between quantum states associated with different parameter choices. The whole machinery of Riemann geometry can then be employed for characterizing phases of quantum matter [16–18]. Moreover, it plays a vital role in understanding equilibrium quantum phase transitions [19,20]. In the thermodynamic limit, phase transitions are signalled by a singularity of the QMT, manifesting as a divergence or a gap. In fact, the topology of the parameter manifold could change across a phase transition [21], leading to a change of the Euler characteristic. The QGT is also useful for characterizing nonequilibrium dynamics, especially close to phase transitions [22–25].

It appears that the QGT can be generalized to the case of NH quantum systems [26]. This generalization is not unique and below we overview several existing choices. When the model is NH, there is no obvious way of defining a proper metric on a submanifold of parameter space (which is a pull-back of a map from the space of states to a ground-state submanifold in parameter space). Indeed, starting from the QGT, one could separate the real and geometric curvature parts in different ways. The complexity of this problem arises from the fact that, generically speaking, we are not dealing with the Hilbert space, but rather with the Krein space, which could have indefinite metric, $J$, such that a proper scalar product of states $\phi$ and $\psi$ is defined as $\langle\phi|J\psi\rangle$ (see, e.g., Refs. [27,28] in the context of $\mathcal{PT}$-symmetric quantum mechanics). This makes a projection to the ground-state parameter submanifold nontrivial.

In this work, we apply quantum geometry to extract the phase diagrams of various NH Su-Schrieffer-Heeger (SSH) models with periodic boundary conditions (PBC). In Sec. II, we present the most general SSH model and introduce the QMT and its NH generalization. We then compute the QMT for various SSH models. We treat these in order of increasing complexity, starting from the Hermitian SSH with real-hopping parameters in Sec. III. In Sec. IV, the pseudo-Hermitian Hamiltonian with real nonreciprocal intracell hoppings is studied. We find that the QMT arising in the NH phases is degenerate, which effectively leads to a dimensional reduction of the associated quantum geometry from two-dimensional to one-dimensional. This phenomenon is not known to occur for Hermitian quantum geometries and appears to be a characteristic of NH ones. Recently, the QGT of a NH system has been experimentally measured [29]. We will show that they have, in fact, observed the dimensional reduction that we predict theoretically here. Section V proceeds with the study of the Hermitian SSH model with complex-hopping amplitudes, and its analogous NH version with the same complex left- and right-handed intracell hopping parameters is analyzed in Sec. VI. The SSH Hamiltonian in Sec. VII is similar to that of Sec. IV, but we now include phases into the nonreciprocal hopping parameters. Again, we find a dimensional reduction of the associated quantum geometry in the NH phases. We note that for all models that we have considered, the phase diagrams known from calculations of the Chern number are faithfully reproduced by the QMT. In Sec. VIII, we discuss the physical implications of our findings. In particular, the aforementioned dimensional reduction implies that there exists a particular choice of parameters for which the associated QMT component is equal to zero. By using NH linear response theory, this means that we can apply a perturbation along this direction in the parameter space, such as a driving or a quench, without exciting the system to a higher energy eigenstate, effectively leaving it unaffected. Last, we discuss a possible application of the pseudo-Riemannian manifolds which arise as the quantum geometries of pseudo-Hermitian quantum systems. Whereas Riemannian spaces only contain space-like and no time-like directions, pseudo-Riemannian manifolds appear as space-time in the context of general relativity. This opens up the possibility of performing 'quantum simulations' of space-time.

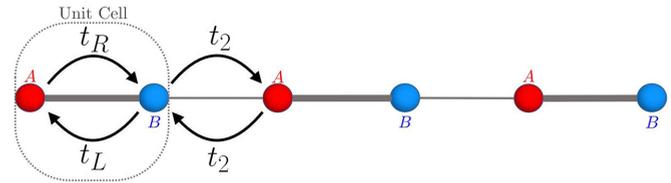

FIG. 1. A general schematic representation of an SSH chain with sublattices *A* and *B*. The hopping parameters are the right-handed intracell $t_R$, the left-handed intracell $t_L$, and the intercell $t_2$.

## II. MODEL AND THEORETICAL FRAMEWORK

### A. NH SSH model

The SSH model is one of the simplest models to describe a topological insulator. It consists of spinless fermions moving on a one-dimensional lattice (i.e., chain) with staggered hopping amplitudes, as shown in Fig. 1. The chain contains $N$ unit cells, each of them with two sites, one on sublattice *A* and another on sublattice *B*. Here, we consider real-valued intercell hopping, $t_2 \in \mathbb{R}$, between nearest unit cells for simplicity. We have checked that the physical results do not change if this quantity is complex.[2] We will be considering different choices for the left- and right-handed intracell hopping amplitudes $t_L$ and $t_R$, which lead to the various SSH models that we consider in the following sections. The Hamiltonian is generally of the following form:

$$\hat{H} = \sum_{i=1}^{N}(t_L\, a_i^\dagger b_i + t_R\, b_i^\dagger a_i) + t_2 \sum_{i=1}^{N-1}(b_i^\dagger a_{i+1} + a_{i+1}^\dagger b_i). \quad (1)$$

---

[1]Superfluid weight is an observable that quantifies the ability of the system to support superfluid transport.

[2]Since the coordinates in the parameter space will involve, for instance, $t_R/t_2$ instead of $t_R$ and $t_2$ separately, only the real- or complex-valuedness of the fraction plays a key role. Therefore, we can choose $t_2 \in \mathbb{R}$ and define $t_L$ and $t_R$ later on.

Here, $a_i^\dagger(a_i)$ denotes the creation (annihilation) operator of a fermionic particle on sublattice $A$ of the $i$th unit cell, and the same for operator $b_i^\dagger(b_i)$ on sublattice $B$. The corresponding Bloch Hamiltonian is

$$H = \begin{pmatrix} 0 & t_R + t_2 e^{-ik} \\ t_L + t_2 e^{ik} & 0 \end{pmatrix}. \quad (2)$$

Its energy eigenvalues are given by

$$E_\pm = \pm\left(t_R t_L + t_2^2 + t_R t_2 e^{ik} + t_L t_2 e^{-ik}\right)^{1/2}, \quad (3)$$

which are generally complex numbers. Now, we define the quantity

$$\varepsilon \equiv \left(\frac{E_\pm}{t_2}\right)^2 = 1 + \frac{t_R}{t_2}\frac{t_L}{t_2} + \frac{t_R}{t_2}e^{ik} + \frac{t_L}{t_2}e^{-ik}, \quad (4)$$

alongside with its complex conjugate $\varepsilon^*$ for convenience. The left- and right-eigenstates of the Bloch Hamiltonian are given by[3]

$$\langle\psi^L| = (|\psi^L\rangle)^\dagger = \frac{1}{\sqrt{2}}\left(\pm 1 \quad \frac{\frac{t_R}{t_2} + e^{-ik}}{\varepsilon^{1/2}}\right),$$
$$\langle\psi^R| = (|\psi^R\rangle)^\dagger = \frac{1}{\sqrt{2}}\left(\pm 1 \quad \frac{\frac{t_L^*}{t_2} + e^{-ik}}{\varepsilon^{*1/2}}\right), \quad (5)$$

where the superscript $*$ denotes the complex conjugation.

There are two ways to extend a Hermitian hopping model to the NH domain. The first is to consider NH complex hopping amplitudes. When the same phase is attributed to both left- and right-handed intracell hoppings, the model is equivalent to a system in which a magnetic flux is piercing each plaquette. The second possibility is to consider instead a real but nonreciprocal term in the original Hermitian intracell hopping amplitude. When all parameters are real, this leads to a preferential direction of propagation for bulk states under PBC [4]. Here, we considered both options to construct different versions of the SSH model, as one will observe in the definition of $t_L$ and $t_R$ later on.

### B. Quantum geometry

While conventional methods to describe topological insulators rely on the momentum space $k$, the geometry of these materials also contains the topological characteristics [21, 30–32] and other interesting features that will be shown in this work.

#### 1. Hermitian quantum metric tensor

Consider a Hermitian Hamiltonian parameterized by a set of degrees of freedom $\lambda \equiv \{\lambda^\mu\}_{\mu=1}^K$, $H = H(\lambda)$. We are only interested in differentiable and normalized eigenstates in this parameter space. Thus, the distance between the states of a Hamiltonian separated by an infinitesimal $d\lambda$ defines the *fidelity*:

$$F(\lambda, \lambda + d\lambda) \equiv |\langle\psi(\lambda)|\psi(\lambda + d\lambda)\rangle|$$
$$= 1 - \tfrac{1}{2}(d\lambda)^2 \chi_F + \cdots, \quad (6)$$

which has the property $0 \leqslant F(\lambda, \lambda + d\lambda) \leqslant 1$ [30]. A Taylor expansion of the fidelity about $d\lambda \to 0$ gives us the *fidelity susceptibility*,

$$\chi_F = \frac{\partial \lambda_\mu}{\partial \lambda}\frac{\partial \lambda_\nu}{\partial \lambda} g_{\mu\nu}, \quad (7)$$

where the vector $\partial\lambda_\mu/\partial\lambda$ denotes the direction of the short displacement $d\lambda$ in the parameter space [30], and $g_{\mu\nu}$ is the symmetric QMT defined as

$$g_{\mu\nu} = \tfrac{1}{2}[\langle\partial_\mu\psi|\partial_\nu\psi\rangle - \langle\partial_\mu\psi|\psi\rangle\langle\psi|\partial_\nu\psi\rangle + (\mu \leftrightarrow \nu)]. \quad (8)$$

In general, away from the quantum critical points, the fidelity vanishes exponentially with the size of a many-body system [32]. Hence, it is more convenient to study a model through the second-derivative term $\chi_F$, which exhibits a maximum or diverges at a critical point [32]. The fidelity susceptibility also follows a scaling law close to a critical point, which can be used to reveal possible topological phase transitions [32].

The square of the fidelity defines a coordinate-independent infinitesimal distance [21]

$$ds^2 = 1 - |\langle\psi(\lambda)|\psi(\lambda + d\lambda)\rangle|^2 \simeq \chi_{\mu\nu} d\lambda^\mu d\lambda^\nu, \quad (9)$$

where the QGT,

$$\chi_{\mu\nu} \equiv \langle\partial_\mu\psi|\partial_\nu\psi\rangle - \langle\partial_\mu\psi|\psi\rangle\langle\psi|\partial_\nu\psi\rangle, \quad (10)$$

is the $\mathcal{O}(d\lambda^2)$-term after the Taylor expansion about $d\lambda \to 0$. Notice that the QMT, Eq. (8), is actually the real and symmetric part of the geometric tensor, Eq. (10),

$$g_{\mu\nu} = \text{Re}[\chi_{\mu\nu}] = \tfrac{1}{2}[\chi_{\mu\nu} + \chi_{\nu\mu}], \quad (11)$$

and it is the key term in Eq. (7). Finally, the imaginary and antisymmetric part of the QGT defines the Berry curvature [21]

$$F_{\mu\nu} = i[\chi_{\mu\nu} - \chi_{\nu\mu}] = \partial_\mu A_\nu - \partial_\nu A_\mu, \quad (12)$$

where the Berry connection is

$$A_\mu = i\langle\psi|\partial_\mu\psi\rangle. \quad (13)$$

Since the Berry formalism can be used to characterize different topological phases and topological phase transitions [4], it is natural to believe that the QMT may also yield some interesting topological properties of a system. Consequently, the latter is the focus of this work.

---

[3]The normalization $\langle\psi^L|\psi^R\rangle = 1$ has been considered here, with the normalization factor $N^{L(R)}(E; \lambda, k) = 1/\sqrt{2}$. Thus, $\langle\psi^{L(R)}|\psi^{L(R)}\rangle \neq 1$. One can always consider the normalization $\langle\psi^{L(R)}|\psi^{L(R)}\rangle = 1$ instead, such as in Ref. [29]. In this case, each normalization factor contains information about both left- and right-eigenstates, and becomes a function involving energy eigenvalues $E$ and parameters $\lambda^\mu$.

Equation (9), together with invariance under the exchange $d\lambda^\mu \longleftrightarrow d\lambda^\nu$, yields to the analog metric equation in general relativity:

$$ds^2 = g_{\mu\nu} d\lambda^\mu d\lambda^\nu. \quad (14)$$

The corresponding covariance property always allows one to locally diagonalize the QMT, such that only the diagonal QMT components,

$$g_{\mu\mu} = \langle \partial_\mu \psi | \partial_\mu \psi \rangle - \langle \partial_\mu \psi | \psi \rangle \langle \psi | \partial_\mu \psi \rangle$$
$$= \langle \partial_\mu \psi | (\mathbb{1} - |\psi\rangle\langle\psi|) | \partial_\mu \psi \rangle, \quad (15)$$

are not equal to zero. Since $g_{\mu\mu}$ can be written as the expectation value in the state $|\partial_\mu \psi\rangle$ of the identity matrix $\mathbb{1}$ minus a projection operator $|\psi\rangle\langle\psi|$, it holds that $g_{\mu\mu} \geqslant 0$. If we exclude coordinates $\lambda^\kappa$ for which $g_{\kappa\kappa} = 0$, we are left with a metric of purely positive signature. For coordinate patches where the metric is regular, the metric signature is constant. We thus conclude that the quantum geometry of Hermitian Hamiltonians is necessarily Riemannian, and this has been the manifold for most of the research thus far [21,30,32–36]. Furthermore, for real $\lambda^\mu$, one can easily verify that the QMT components are real as well.

### 2. Non-Hermitian quantum metric tensor

In the case of NH systems, it is natural to think about a NH version of the QMT. First we introduce the NH version of the QGT tensor, given by

$$\chi^{\alpha\beta}_{\mu\nu} \equiv \langle \partial_\mu \psi^\alpha | \partial_\nu \psi^\beta \rangle - \langle \partial_\mu \psi^\alpha | \psi^\beta \rangle \langle \psi^\alpha | \partial_\nu \psi^\beta \rangle, \quad (16)$$

with the orthonormality $\langle \psi^\alpha | \psi^\beta \rangle = 1 \; \forall \alpha, \beta \in \{L, R\}$, specifying the left and right vector spaces of the eigenfunction $\psi$.

While for the Hermitian QMT the real and symmetric part of the QGT coincide, this is not necessarily the case for NH systems. This introduces an ambiguity, which allows for three different definitions of the QMT to be taken:

(1) the QMT is the symmetric part of the QGT with respect to the indices $\mu$ and $\nu$,

$$g^{\alpha\beta}_{\mu\nu} = \frac{\chi^{\alpha\beta}_{\mu\nu} + \chi^{\alpha\beta}_{\nu\mu}}{2}, \quad (17)$$

(2) the QMT is the real part of the QGT,

$$\text{Re}\big[\chi^{\alpha\beta}_{\mu\nu}\big] = \frac{\chi^{\alpha\beta}_{\mu\nu} + \chi^{\beta\alpha}_{\nu\mu}}{2}, \quad (18)$$

(3) the QMT is the real and symmetric part of the QGT,

$$G_{\mu\nu} = \text{Re}\big[G^{\alpha\beta}_{\mu\nu}\big] = \frac{\chi^{\alpha\beta}_{\mu\nu} + \chi^{\beta\alpha}_{\nu\mu} + \chi^{\alpha\beta}_{\nu\mu} + \chi^{\beta\alpha}_{\mu\nu}}{4}. \quad (19)$$

The above definitions are equivalent for Hermitian and pseudo-Hermitian systems, such as the model with real non-reciprocal hopping that we will consider in Sec. IV. However, they differ for general NH models, such as those studied in Secs. VI and VII.

There has been much confusion in the literature about which should be the correct way to define the QMT for NH systems. Different authors use different definitions, and often do not specify which definition they are using. For example, Refs. [15,37] only mention that they take the real part without specifying symmetrization [which could mean that they are using Eq. (18)], while Refs. [18,38] emphasize that their QMT is both real and symmetrized [Eq. (19)]. We tried to compile all possibilities, to clarify the options and compare them in our work. Equation (17) takes into account the left-right and index symmetries when changing parameters/coordinates. Equation (18) relies on the reality of the geometric tensor. These two versions are less stringent than Eq. (19), which forces the space to be symmetric and real. The minimal requirement for a valid metric tensor is to be symmetric in its 'dimensional' indices, that is $\mu$ and $\nu$ in our notation. There is no *a priori* physical reason for an additional symmetrization in $\alpha$ and $\beta$ to define a metric tensor that is always real. As such, we opted not to perform this additional symmetrization initially to not place unnecessary constraints on the quantum metric tensor, but to discuss the implications of the different choices for the final results. Further, complex metrics naturally appear in the context of complex geometry [39], so there is no *a priori* mathematical reason to demand that the metric is always real either.

The energy eigenvalues of a NH Hamiltonian can be complex even when the space parameters are real-valued, and the same holds for the corresponding eigenfunctions. Therefore, Eq. (17) is not restricted to the Riemannian manifold anymore. In terms of symmetry of the above QMT, nothing changes when exchanging $\lambda^\mu \leftrightarrow \lambda^\nu$, and an additional complex-conjugated relation arises when interchanging $\alpha \leftrightarrow \beta$:

$$g^{\alpha\beta}_{\mu\nu} = g^{\alpha\beta}_{\nu\mu}, \quad g^{\alpha\beta}_{\mu\nu} = \big(g^{\beta\alpha}_{\mu^*\nu^*}\big)^*. \quad (20)$$

In this work, eigenstates of the Bloch Hamiltonian, Eq. (5), will be applied in Eq. (17). Later, we eliminate the momentum dependence of the QMT by summing it over the entire Brillouin zone (BZ) and considering the factor $1/N$ [21]. This method results in an intensive metric that completely captures the topological phases and topological phase transitions:

$$g^{\alpha\beta}_{\mu\nu}(\lambda) = \frac{1}{N} \sum_{k \in BZ} g^{\alpha\beta}_{\mu\nu}(\lambda; k) \xrightarrow{N \to \infty} \frac{1}{2\pi} \int_{BZ} dk\, g^{\alpha\beta}_{\mu\nu}(\lambda; k),$$

$$g^{\alpha\beta}_{\mu\nu}(\lambda) \xrightarrow{\text{notation}} g^{\alpha\beta}_{\mu\nu}. \quad (21)$$

For a continuous momentum space, the metric is also continuous. This can be achieved by considering the thermodynamic limit ($N \to \infty$). Then, the sum turns into a Riemann integral. Since both $g^{\alpha\beta}_{\mu\nu}(\lambda; k)$ and $g^{\alpha\beta}_{\mu\nu}(\lambda)$ satisfy Eq. (20), we keep the notation $g^{\alpha\beta}_{\mu\nu}$ in this work for simplicity.

Note that the partial derivative operation $\partial_\mu$ in Eq. (16) yields the same QMT for $\pm$ eigenstates in Eq. (5) because $\partial_\mu(\pm 1) = 0$.

Finally, it is worth mentioning that the orthonormality condition in Eq. (16), when using eigenstates of a Hamiltonian like Eq. (5), is satisfied only when we combine $L$ and $R$ terms, such as $\langle \psi^L | \psi^R \rangle = 1$. Since the QMT is only well-defined for orthonormal bases, $g^{\alpha\alpha}_{\mu\nu}$ with $\alpha \in \{L, R\}$ are not strictly speaking legitimate QMT's. Nevertheless, as we will show later, it is interesting to compute them, and we will not distinguish them from the real QMT in terms of notations.

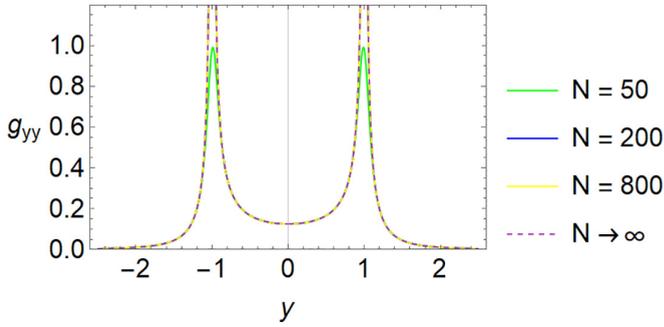

FIG. 2. Metric tensor $g_{yy}$ in the parameter space $y$ for $N = \{50, 200, 800\}$ and $N \to \infty$.

## III. QUANTUM GEOMETRY FOR THE HERMITIAN SSH MODEL: REAL HOPPING (H-SSH)

We start with the most familiar and simplest Hermitian SSH model with real intracell hopping parameter $t_R = t_L \equiv t \in \mathbb{R}$ [12] to introduce the concept of QMT. The eigenstates are given by

$$\langle \psi| = (|\psi\rangle)^\dagger = \frac{1}{\sqrt{2}}\left(\pm 1 \quad \frac{y+e^{-ik}}{\varepsilon_r^{1/2}}\right), \quad (22)$$

and Eq. (4) becomes $\varepsilon_r = 1 + y^2 + 2y\cos(k) \in \mathbb{R}$, with $y \equiv t/t_2 \in \mathbb{R}$ the coordinate of the parameter space and the subscript $r$ denotes real intracell hopping amplitude.

By applying Eqs. (17)–(21), one obtains the one-component QMT

$$g_{yy} = \frac{1}{N}\sum_{k\in BZ}\frac{\sin^2(k)}{4\,\varepsilon_r^2} \xrightarrow{N\to\infty} \begin{cases} \frac{1}{8y^2(y^2-1)} & |y| > 1, \\ \frac{1}{8(1-y^2)} & |y| < 1. \end{cases} \quad (23)$$

Note that the singularity manifests as divergences for $y^2 = 1$, and that $g_{yy}$ is a positive quantity, in agreement with the Hermitian definition of the QMT detailed in Sec. II B.

We plot $g_{yy}$ in Fig. 2. The graphical representations for $N = \{50, 200, 800\}$ and $N \to \infty$ overlap with each other, except at points near the peaks. These maxima become higher when $N$ becomes larger, until turning into divergences in the thermodynamic limit. Here, it is clear that the system reveals the same topological information for almost any system size $N$.

From previous works, we know beforehand that the H-SSH model has topological phase transition points at $|y| = 1$, which separate the trivial phase $|y| > 1$ and the topological regime $|y| < 1$ [4,12]. This agrees with the results of the QMT in Eq. (23) and Fig. 2: the topological phase transitions manifest as a divergent singularity at $|y| = 1$ and each phase has its own metric in the parameter space. Physically, the QMT measures the distance between different states. Since the trivial phase has identical states, the metric tends to zero rapidly for $|y| > 1$ (see Fig. 2). This phase is adiabatically connected to the phase corresponding to $y \to \infty$, for which $g_{yy}$ is identically zero. The topological phase has a fundamental difference between bulk and boundary states, so the metric is bounded below for $|y| < 1$ with a finite value $g_{yy} = 1/8$.

Moreover, the two limits,

$$\lim_{y\to\infty} g_{yy} = 0 \quad \text{and} \quad \lim_{y\to 0} g_{yy} = 1/8,$$

are exactly the two fully dimerized cases of the usual SSH model [4]: $y \to \infty$ and $y \to 0$ belong to the trivial and topological phases, respectively. A topological regime is unaltered under an adiabatic deformation up to a phase transition, so $|y| > 1$ and $|y| < 1$ are the same phases as $y \to \infty$ and $y \to 0$, respectively.

The above results ensure the validity of this approach when describing topological insulators.

## IV. PSEUDO-HERMITIAN SSH MODEL: REAL NONRECIPROCAL HOPPINGS (NH-SSH-NR)

Continuing with real hopping parameters, we now consider an additional nonreciprocal term $\delta$ as $t_R \equiv t - \delta \in \mathbb{R}$ and $t_L \equiv t + \delta \in \mathbb{R}$, which introduces non-Hermiticity to the model. Defining $\{y \equiv t/t_2, z \equiv \delta/t_2\} \in \mathbb{R}$, the eigenstates in Eq. (5) are

$$\langle \psi^L| = (|\psi^L\rangle)^\dagger = \frac{1}{\sqrt{2}}\left(\pm 1 \quad \frac{y-z+e^{-ik}}{\varepsilon_{\mathrm{NRr}}^{1/2}}\right),$$

$$\langle \psi^R| = (|\psi^R\rangle)^\dagger = \frac{1}{\sqrt{2}}\left(\pm 1 \quad \frac{y+z+e^{-ik}}{\varepsilon_{\mathrm{NRr}}^{*\,1/2}}\right), \quad (24)$$

where $\varepsilon_{\mathrm{NRr}} = 1 + y^2 - z^2 + 2y\cos(k) - i2z\sin(k) \in \mathbb{C}$, and the subscript NRr denotes nonreciprocal real intracell hopping amplitudes. By applying Eqs. (17)–(21), the discrete version of all independent QMT components are[4,5]

$$g_{yy}^{LR} = \frac{1}{N}\sum_{k\in BZ}\frac{-[z+i\sin(k)]^2}{4\,\varepsilon_{\mathrm{NRr}}^2} \in \mathbb{R},$$

$$g_{zz}^{LR} = -\frac{1}{N}\sum_{k\in BZ}\frac{[y+\cos(k)]^2}{4\,\varepsilon_{\mathrm{NRr}}^2} \in \mathbb{R},$$

$$g_{yz}^{LR} = \frac{1}{N}\sum_{k\in BZ}\frac{[y+\cos(k)][z+i\sin(k)]}{4\,\varepsilon_{\mathrm{NRr}}^2} \in \mathbb{R}, \quad (25)$$

$$g_{yy}^{\alpha\alpha} = \frac{1}{N}\sum_{k\in BZ}[z^2+\sin^2(k)]K^{\alpha\alpha} \in \mathbb{R},$$

$$g_{zz}^{\alpha\alpha} = \frac{1}{N}\sum_{k\in BZ}[y+\cos(k)]^2 K^{\alpha\alpha} \in \mathbb{R},$$

$$g_{yz}^{\alpha\alpha} = \frac{1}{N}\sum_{k\in BZ}[z(y+\cos(k))]K^{\alpha\alpha} \in \mathbb{R},$$

---

[4]Equations are written in a format such that one can correctly obtain their graphical results by using *Wolfram Mathematica*. Other formats may lead to inconsistent results. For *LR*-QMT components, they are also the appropriate format for computing their Riemann integral using the same program. In addition, the integration can be analytically performed by applying the residue theorem, when the momentum $k$ is written in the exponential form $e^{\pm ik}$.

[5]Keep in mind that $g_{\mu\nu}^{\alpha\alpha}$ with $\alpha \in \{L, R\}$ are not legitimate QMT's because the eigenstates are not normalized, $\langle \psi^\alpha|\psi^\alpha\rangle \neq 1$, and we are considering only PBC.

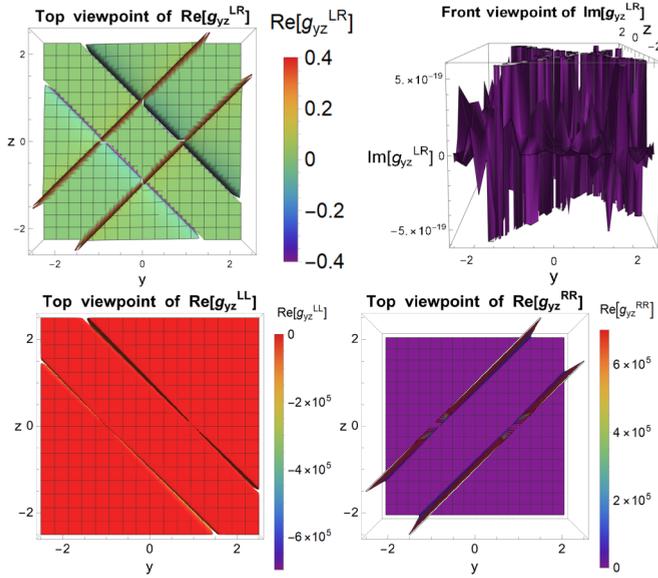

FIG. 3. The QMT components $g_{yz}^{\alpha\beta}$ with $\alpha, \beta \in \{L, R\}$ in the parameter space $y, z \in \mathbb{R}$ for $N = 200$. Their imaginary parts are nonzero (order $10^{-19}$) due to finite-size effects: $\text{Im}[g_{\mu\nu}^{\alpha\beta}] \approx 0 \ \forall \lambda^\mu, \lambda^\nu \in \{y, z\}$. *Note*: keep in mind that $g_{\mu\nu}^{\alpha\alpha}$ with $\alpha \in \{L, R\}$ are not legitimate QMT components.

where the superscript $\alpha \in \{L, R\}$ and

$$K^{LL} = \frac{1}{2\,\varepsilon_-^{1/2}\,\varepsilon_+^{3/2}} - \frac{1}{4\,\varepsilon_+^2} \ \in \mathbb{R},$$

$$K^{RR} = \frac{1}{2\,\varepsilon_-^{3/2}\,\varepsilon_+^{1/2}} - \frac{1}{4\,\varepsilon_-^2} \ \in \mathbb{R}, \quad (26)$$

$$\varepsilon_\pm \equiv 1 + (y \pm z)^2 + 2(y \pm z)\cos(k) \ \in \mathbb{R}.$$

The singular points of a QMT component are those with a vanishing denominator and a finite numerator. Therefore, the previous equations imply that the metric diverges at

$$\varepsilon_{\text{NRr}} = 0 \implies z = \pm y \pm 1,$$
$$\varepsilon_- = 0 \implies z = y \pm 1, \quad (27)$$
$$\varepsilon_+ = 0 \implies z = -y \pm 1.$$

Specifically, the QMT components with *LR*-combinations (*LR*-QMT components) have four topological phase transition lines, given by $z = \pm y \pm 1$. However, *LL*-QMT components and *RR*-QMT components only contain two: $z = y \pm 1$ and $z = -y \pm 1$, respectively. This phenomenon can be clearly visualized in Fig. 3, where the $yz$-component for $N = 200$ is presented as an example (see Appendix A).

This demonstrates that the *LL*- and *RR*-QMT components only provide half of the total phase diagram, while the *LR*-QMT components contain the complete information. Thus, it confirms the need to use biorthogonal quantum mechanics when dealing with NH systems [21,40]. Furthermore, the imaginary part of the *LR*-QMT components is effectively zero. The latter also occurs for the *LL*- and *RR*-QMT components.

We now take $N \to \infty$, where the sum for the QMT components turns into an integral, which can be solved exactly. The 3D top-view representations of $g_{yy}^{LR}$, $g_{zz}^{LR}$, and $g_{yz}^{LR}$ are shown in Fig. 4. We see that they faithfully reproduce the boundaries of the topological phase diagram presented in Fig. 5, where the pairs of numbers taking values 0 and 1/2 are the NH topological winding numbers [12,41]. The *green* and *red* areas in Fig. 5 describe the trivial and topological phases, respectively, while the *blue* and *yellow* regions depict NH topological phases. Finally, the diagonal QMT components in Fig. 4 show two distinct geometries in the *green* domain with the same two-components winding number (0,0) (see Fig. 5). The regions I and II are not adiabatically connected due to the phase transitions. If the two regions would be exactly the same topological regime, then the QMT components should be symmetric upon exchanging $y \leftrightarrow z$, as it happens for the QMT components in the *blue* and *yellow* domains. However, this is not the case. Hence, regions I and II should be different. This distinction also manifests when one graphically depicts the winding vector around the two exceptional points [41,42] (see Appendix A). Region I is the area in which the exceptional points are positioned completely to the right or to the left of the winding vector. In contrast, in region II, the winding vector is in between the two exceptional points, encircling neither of them. Mathematically, these exceptional points are given by the imaginary parts of the components of the Bloch Hamiltonian in the Pauli matrix basis, $(\pm\text{Im}[d_y(k)], \mp\text{Im}[d_x(k)])$, where $d_x(k) = t + t_2\cos(k)$, and $d_y(k) = t_2\sin(k) - i\delta$, such that $H(k) = \sum_i d_i(k) \cdot \sigma_i$ [42]. For this model, these two regions are not connected because as $d_x(k)$ is entirely real, the exceptional points lie along the horizontal axis and $d_y(k)$ does not have any degree of freedom that allows for translations in the real plane. Thus, one must move through one of the exceptional points to be able to access the other region. Since the two trivial phases are not adiabatically connected, they are not the same. From the above discussion, we are now denoting this region II as NH trivial because it only occurs at finite $\delta$, while the Hermitian phase I includes $\delta = 0$.

Here, it is worth pointing out that the symmetry by exchanging $\lambda^\mu \to -\lambda^\mu$ relies on the metric equation $ds^2 = g_{\mu\nu}^{LR} d\lambda^\mu d\lambda^\nu$, where the line element also has the negative sign, instead of the QMT itself.

The QMT's corresponding to distinct regions in Fig. 5 provide a more complete understanding of topological systems. All components are real-valued, which corresponds to the conventional quantum geometry framework. Moreover, we note the existence of negative diagonal QMT components in all topological regions, which is not possible for any Hermitian system. Therefore, the extension from the Riemannian to the pseudo-Riemannian manifold definitely opens the door to a more complex realm.

The next point is about the topological phase transition lines. These curves are coincidentally lightlike trajectories when the metric equation equals zero, $ds^2 = 0$. One can check that there is an exception in the trivial domain due to the distinction between regions I and II discussed above. It is interesting to observe from Fig. 5 that mathematically

$$g_{\mu\nu}^{LR}(\text{"green"}) + g_{\mu\nu}^{LR}(\text{"red"}) = g_{\mu\nu}^{LR}(\text{"blue"}) + g_{\mu\nu}^{LR}(\text{"yellow"}).$$

However, it is not clear, from a physics perspective, why this relation holds.

Finally, the degeneracy of the QMT in the *blue* and *yellow* areas gives rise to a dimensional reduction. This can be clearly appreciated when we perform a transformation of the QMT, $g_{\mu\nu} = \frac{\partial \lambda^\delta}{\partial \lambda^\mu} \frac{\partial \lambda^\sigma}{\partial \lambda^\nu} g_{\delta\sigma}$, to the new coordinates $x_\pm \equiv y \pm z \in \mathbb{R}$:

$$\blacksquare \quad g_{\mu\nu}^{LR} = \begin{pmatrix} -\frac{1}{16x_-^2} & 0 \\ 0 & 0 \end{pmatrix},$$

$$\blacksquare \quad g_{\mu\nu}^{LR} = \begin{pmatrix} 0 & 0 \\ 0 & -\frac{1}{16x_+^2} \end{pmatrix},$$

$$\blacksquare \quad g_{\mu\nu}^{LR} = \begin{pmatrix} 0 & \frac{1}{16(1-x_-x_+)} \\ \frac{1}{16(1-x_-x_+)} & 0 \end{pmatrix},$$

$$\blacksquare \quad g_{\mu\nu}^{LR} = \begin{pmatrix} -\frac{1}{16x_-^2} & -\frac{1}{16(1-x_-x_+)} \\ -\frac{1}{16(1-x_-x_+)} & -\frac{1}{16x_+^2} \end{pmatrix}. \quad (28)$$

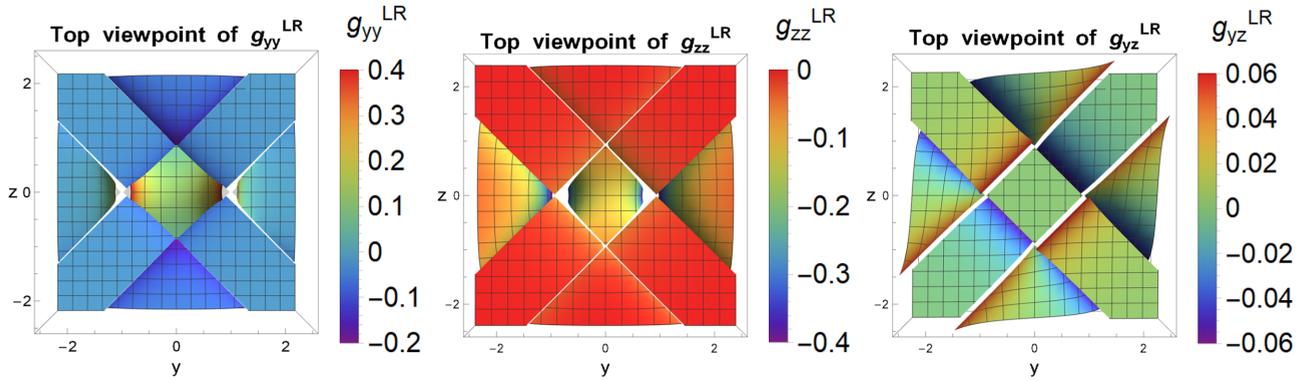

FIG. 4. QMT components Re[$g_{\mu\nu}^{LR}$] with $x^\mu, x^\nu \in \{y, z\}$ in the parameter space $y, z \in \mathbb{R}$ for $N \longrightarrow \infty$. Top viewpoint. Their imaginary parts are zero: Im[$g_{\mu\nu}^{LR}$] = 0 $\forall \lambda^\mu, \lambda^\nu \in \{y, z\}$. Therefore, $g_{\mu\nu}^{LR} = \text{Re}[g_{\mu\nu}^{LR}]$.

Equation (28) implies that the *blue* regions have a dark coordinate $x_+$, whereas this happens for the $x_-$-direction in the *yellow* areas.

This dimensional reduction could be a generic phenomenon in NH models, but a mathematical proof of this statement demands further investigation. It requires a model which has at least two parameters driving the phase transitions, and the metric tensor must be singular and/or degenerate. In our NH-SSH-NR case, $y = t/t_2$ and $z = \delta/t_2$.

In usual two-particle problems, the system can be described in terms of the generalized "center-of-mass $(y + z)$" and "relative coordinates $(y - z)$". If one of those is zero, the two parameters are equal or are the negative of each other, which is equivalent because the phase transitions occur for both positive and negative values of the parameters (i.e., the phase diagram in Fig. 5 is symmetric). Therefore, if the parameters driving the phase transition are reduced from two to one, then there is a dimensional reduction. At the dimensional reduction, the two parameters driving the phase transition synchronize (in phase or out-of-phase) and the intracell hopping becomes unidirectional (either $t + \delta = 0$ or $t - \delta = 0$). An inspection of the phase diagram shown in Fig. 5 indicates that along the diagonals $y = z$ or $y = -z$, it is only possible to probe transitions from the NH to the Hermitian topological phases. Therefore, the dimensional reduction observed here requires a NH topological system.

## V. HERMITIAN SSH MODEL: COMPLEX HOPPING (H-SSH-C)

We now introduce a complex hopping parameter into the Hermitian SSH model: $t_R \equiv te^{i\theta} \in \mathbb{C}$ and $t_L \equiv te^{-i\theta} \in \mathbb{C}$, with $t \in \mathbb{R}$ and $\theta \in [0, 2\pi)$. Following the same method as before, the two complex parameters, $\{u \equiv (t/t_2)e^{i\theta}, u^* \equiv$ $(t/t_2)e^{-i\theta}\} \in \mathbb{C}$, give rise to three independent QMT components: $\{g_{uu} \in \mathbb{C}, g_{u^*u^*} \in \mathbb{C}, g_{uu^*} \in \mathbb{R}\}$. In the thermodynamic limit, the QMT reads

$$g_{\mu\nu} = \begin{cases} \begin{pmatrix} -\frac{1}{16u^2} & -\frac{1}{16(1-uu^*)} \\ -\frac{1}{16(1-uu^*)} & -\frac{1}{16u^{*2}} \end{pmatrix} & |u|^2 > 1, \\ \begin{pmatrix} 0 & \frac{1}{16(1-uu^*)} \\ \frac{1}{16(1-uu^*)} & 0 \end{pmatrix} & |u|^2 < 1. \end{cases} \quad (29)$$

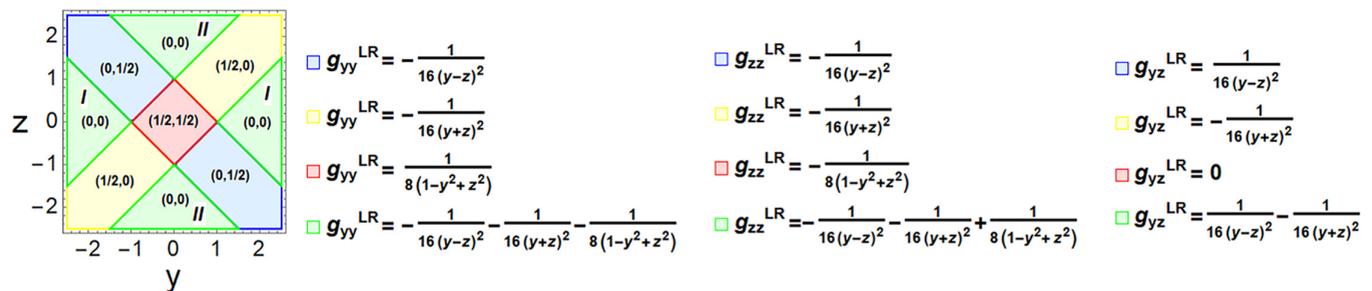

FIG. 5. The phase diagram with associated metric components $g_{yy}^{LR}$, $g_{zz}^{LR}$, and $g_{yz}^{LR}$ in the parameter space $y, z \in \mathbb{R}$. The symmetry by exchanging $x^\mu \to -x^\mu$ relies on the metric equation $ds^2 = g_{\mu\nu}^{LR} d\lambda^\mu d\lambda^\nu$, where the line element also has the negative sign, instead of the QMT itself. This domain matches perfectly with the topological phase diagram found by using a more conventional method in Refs. [12,41]. Different topological phases are distinguished by colours and the two-component winding number $(\nu_1, \nu_2)$. Hermitian topological phases have $\nu_1 = \nu_2$, while $\nu_1 \neq \nu_2$ only happens for NH regimes. Specifically, (0,0) represents a Hermitian and NH trivial phase (I and II, respectively), while (1/2, 1/2) is a Hermitian topological phase, and (0, 1/2) and (1/2, 0) represent NH topological phases. Notice that the Hermitian phase diagram is recovered when $z = 0$.

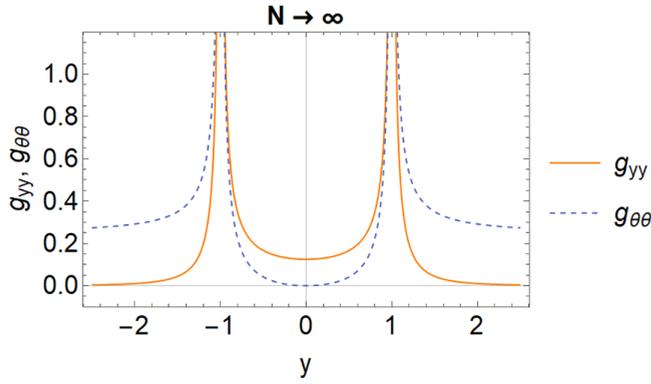

FIG. 6. $g_{yy}$ and $g_{\theta\theta}$ in the parameter space $\{y \equiv t/t_2, \theta\}$ for $N \to \infty$ (i.e., thermodynamic limit). Note that both QMT components are independent of the parameter $\theta$.

We see that the metric is singular at $|u|^2 = 1$, dividing the domain into two: $|u|^2 > 1$ and $|u|^2 < 1$. Note that $g_{uu}$ and $g_{u^*u^*}$ vanish for $|u|^2 < 1$, and the QMT components are purely real-valued in this region. We now switch to variables $\{y, \theta\}$:

$$g_{\mu\nu} = \begin{cases} \begin{pmatrix} \frac{1}{8y^2(y^2-1)} & 0 \\ 0 & \frac{2y^2-1}{8(y^2-1)} \end{pmatrix} & y^2 > 1, \\ \begin{pmatrix} \frac{1}{8(1-y^2)} & 0 \\ 0 & \frac{y^2}{8(1-y^2)} \end{pmatrix} & y^2 < 1, \end{cases} \quad (30)$$

where $g_{y\theta} = g_{\theta y} = 0$, and the metric is diagonal, with real-valued and positive-definite QMT components. That is, we indeed get a Riemannian metric for this Hermitian SSH model, as expected. The topological phase transition occurs at $|u|^2 = y^2 = 1$, splitting the parameter space into the topological ($y^2 < 1$) and trivial ($y^2 > 1$) phases. Looking back at Eq. (29), this argument implies that both the real and imaginary parts of the QMT present information about topological phases and topological phase transitions. Returning to Eq. (30), on one hand, it is clear that the QMT only depends on the parameter $y$, meaning that the topological phenomenon is independent of the phase $\theta$ in the hopping. On the other hand, while the QMT component $g_{yy}$ agrees with the one-component QMT of the H-SSH model in Sec. III, the QMT component $g_{\theta\theta}$ is indeed caused by the complex extension of the hopping amplitudes. The latter gives us some additional information about the system, which is not captured by using conventional methods [4]. This can be seen in Fig. 6, where $g_{\theta\theta}$ behaves reversed with respect to $g_{yy}$. Specifically, $g_{\theta\theta}$ is finite in the trivial regime (i.e., $\lim_{y^2 \to \infty} g_{\theta\theta} = 1/4$) and it goes to zero rapidly in the topological phase (i.e., $\lim_{y^2 \to 0} g_{\theta\theta} = 0$), whereas $\lim_{y^2 \to 0} g_{yy} = 1/8$ in the topological phase and goes to zero in the trivial one.

## VI. NON-HERMITIAN SSH MODEL: COMPLEX HOPPINGS (NH-SSH-C)

Another way to construct a NH SSH model is to have the same complex left- and right-handed intracell hopping amplitudes: $t_R = t_L \equiv te^{i\theta} \in \mathbb{C}$, with $t \in \mathbb{R}$ and $\theta \in [0, 2\pi)$. While $\{u, u^*\} \in \mathbb{C}$ are the two parameters of the system, there is only one nonzero QMT component for each independent $\{L, R\}$-combination of the QMT:

$$g_{uu}^{LR} = \frac{1}{N} \sum_{k \in BZ} \frac{\sin^2(k)}{4 \varepsilon_{NHc}^2} \in \mathbb{C},$$

$$g_{uu^*}^{LL} = \frac{1}{N} \sum_{k \in BZ} \left[ \frac{\sin^2(k)}{4\sqrt{\varepsilon_{NHc}}\sqrt{\varepsilon_{NHc}^*}(u+e^{ik})(u^*+e^{-ik})} \right.$$
$$\left. - \frac{\sin^2(k)}{8(u+e^{ik})^2(u^*+e^{-ik})^2} \right] \in \mathbb{R}, \quad (31)$$

$$g_{uu^*}^{RR} = \frac{1}{N} \sum_{k \in BZ} \left[ \frac{\sin^2(k)}{4\sqrt{\varepsilon_{NHc}}\sqrt{\varepsilon_{NHc}^*}(u^*+e^{+ik})(u+e^{-ik})} \right.$$
$$\left. - \frac{\sin^2(k)}{8 \varepsilon_{NHc} \varepsilon_{NHc}^*} \right] \in \mathbb{R},$$

where $\varepsilon_{NHc} = 1 + u^2 + 2u\cos(k) \in \mathbb{C}$. Like the NH-SSH-NR model in Sec. IV, both $LL$- and $RR$-QMT's cover half of the total topological information (see Appendix B). Thus, one can focus the study on the $LR$-QMT only. Writing the parameter in the complex form $u = u_R + iu_I$, we realize that all QMT components depend only on $g_{uu}^{LR}$:

$$g_{u_R u_R}^{LR} = g_{uu}^{LR},$$
$$g_{u_I u_I}^{LR} = -g_{uu}^{LR}, \quad (32)$$
$$g_{u_R u_I}^{LR} = ig_{uu}^{LR}.$$

Notice that $g_{uu}^{LR}$ in Eq. (31) is exactly the complex version of Eq. (23), where we now have the parameter $u \equiv ye^{i\theta}$ instead of $y$. Therefore, its Riemann integral in the thermodynamic limit also follows this analogy. In the $\{u, u^*\}$-space, the matrix form of the QMT,

$$g_{\mu\nu}^{LR} = \begin{cases} \begin{pmatrix} \frac{1}{8u^2(u^2-1)} & 0 \\ 0 & 0 \end{pmatrix} & |u|^2 > 1, \\ \begin{pmatrix} \frac{1}{8(1-u^2)} & 0 \\ 0 & 0 \end{pmatrix} & |u|^2 < 1, \end{cases} \quad (33)$$

shows clearly a dimensional reduction in the $u^*$-direction. In particular, there is a topological phase transition unit circle in the $\{u_R, u_I\}$-space, $|u|^2 = 1$, that separates the extended topological ($|u|^2 < 1$) and trivial ($|u|^2 > 1$) phases, as shown in Fig. 7. Here, one can observe that both the real and imaginary part of $g_{uu}^{LR}$ contain topological information, as in the previous model. If, however, we consider the third definition of the QMT (i.e., Eq. (19)), which merely extracts the real part of $g_{\mu\nu}^{LR}$, we find that $G_{u^*u^*}$ is nonzero and that there is no dimensional reduction.

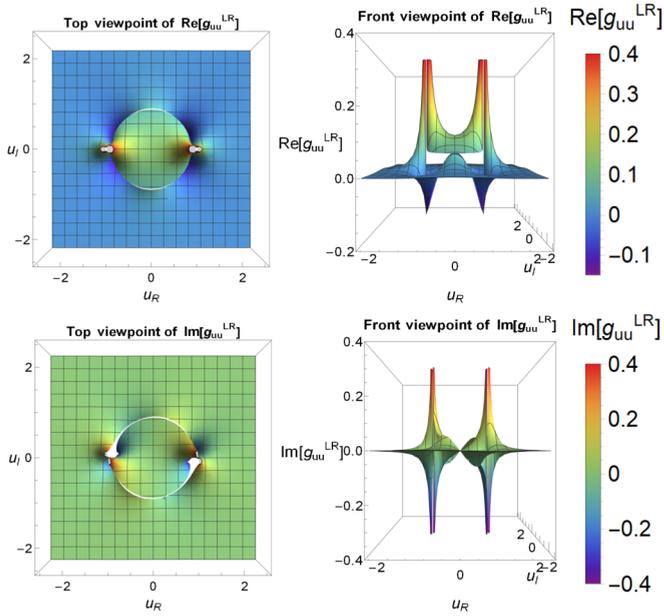

FIG. 7. Real and imaginary parts of the QMT component $g_{uu}^{LR}$ in the parameter space $\{u_R, u_I\} \in \mathbb{R}$ for $N \to \infty$.

## VII. NON-HERMITIAN SSH MODEL: COMPLEX NONRECIPROCAL HOPPING (NH-SSH-NR-C)

We now include phases into the nonreciprocal hopping parameters of the pseudo-Hermitian SSH model presented in Sec. IV, such that $t_R \equiv te^{i\theta} - \delta e^{i\phi} \in \mathbb{C}$ and $t_L \equiv te^{-i\theta} + \delta e^{-i\phi} \in \mathbb{C}$, with $t, \delta \in \mathbb{R}$ and $\theta, \phi \in [0, 2\pi)$ in the principal branch. Thus, the parameters of the model are again $\{u, u^*\} \in \mathbb{C}$ together with $\{v \equiv (\delta/t_2)e^{i\phi}, v^* \equiv (\delta/t_2)e^{-i\phi}\} \in \mathbb{C}$. In this case, we find that the most promising *LR*-combination results in three independent QMT components, due to the following additional relations:

$$\begin{aligned}
g_{uu}^{LR} &= g_{vv}^{LR} = -g_{uv}^{LR}, \\
g_{u^*u^*}^{LR} &= g_{v^*v^*}^{LR} = g_{u^*v^*}^{LR}, \\
g_{uu^*}^{LR} &= g_{uv^*}^{LR} = -g_{vv^*}^{LR} = -g_{u^*v}^{LR}.
\end{aligned} \quad (34)$$

Here, we choose the set of $g_{\{u,u^*\}}^{LR}$ QMT components for simplicity. In the thermodynamic limit, the matrix representation of the metric tensor,

$$g_{\mu\nu}^{LR} = \begin{pmatrix} g_{uu}^{LR} & g_{uu^*}^{LR} \\ g_{uu^*}^{LR} & g_{u^*u^*}^{LR} \end{pmatrix},$$

for distinct topological regions reads

$$g_{\mu\nu}^{LR} = \begin{cases}
\begin{pmatrix} -\frac{1}{16(u-v)^2} & 0 \\ 0 & 0 \end{pmatrix} & |u-v|^2 > 1 \wedge |u^*+v^*|^2 < 1, \\
\begin{pmatrix} 0 & 0 \\ 0 & -\frac{1}{16(u^*+v^*)^2} \end{pmatrix} & |u-v|^2 < 1 \wedge |u^*+v^*|^2 > 1, \\
\begin{pmatrix} 0 & \frac{1}{16}\frac{1}{1-(u-v)(u^*+v^*)} \\ \frac{1}{16}\frac{1}{1-(u-v)(u^*+v^*)} & 0 \end{pmatrix} & |u-v|^2 < 1 \wedge |u^*+v^*|^2 < 1, \\
\begin{pmatrix} -\frac{1}{16(u-v)^2} & -\frac{1}{16}\frac{1}{1-(u-v)(u^*+v^*)} \\ -\frac{1}{16}\frac{1}{1-(u-v)(u^*+v^*)} & -\frac{1}{16(u^*+v^*)^2} \end{pmatrix} & |u-v|^2 > 1 \wedge |u^*+v^*|^2 > 1,
\end{cases} \quad (35)$$

where $\wedge$ denotes the logic *and*. One may verify that the above QMT components reduce to those of the H-SSH-C model in Sec. V [i.e., Eq. (29)] when $v = 0 = v^*$. Furthermore, the QMT components of the NH-SSH-NR model in Sec. IV can be obtained with $\theta = 0 = \phi$. Last, the mathematical structure of Eqs. (28) and (35) are the same, which allows one to directly identify the QMT corresponding to the distinct topological phases: going from up to down in Eq. (35), they are "NH topological phase," "NH topological phase," "Hermitian topological phase," and "trivial phases." The phase transition curves of the system are $|u-v|^2 = 1$ and $|u^*+v^*|^2 = 1$. The above analysis implies that the current SSH model combines all the physical results in both NH-SSH-NR and H-SSH-C cases. Note that while this metric has complex components and dimensional reduction, when we use the definition of the QMT that is both real and symmetric [Eq. (19)], the metric becomes real and we no longer have dimensional reduction, as we saw in Sec. VI.

## VIII. PHYSICAL IMPLICATIONS

### A. Signature of the quantum metric and dimensional reduction

We found that when we consider the NH version of the QMT in Eq. (17), we can have negative QMT components. Choosing the definition of the metric such that it is both real and symmetric [i.e., Eq. (19)], nonetheless, makes sense because of its function as a measure of distances on the manifold.[6] This real symmetric QMT is also the metric that has been derived for $\mathcal{PT}$-symmetric Hamiltonians specifically [38]. Still, NH QMT's significantly increase the scope of possible QMT's from Riemannian to pseudo-Riemannian (with mixed positive and negative signatures). In the context of relativity, both special and general, one typically considers pseudo-Riemannian manifolds. Here, the coordinates with negative signature describe time-like dimensions in the parameter manifold. We believe that these observations are general features of NH quantum systems, going beyond the scope of NH SSH models.

---
[6]Even so, in Ref. [39], the author works with complex QMT components in complex manifolds.

We also observe that a dimensional reduction of the parameter space appears in NH phases. We remind the reader that the signature of a metric is fixed in regions where the metric is regular. Therefore, dimensional reduction from 2D to 1D only occurs at the loci where the metric is not regular. These are precisely the critical points where a phase transition occurs. The occurrence of dimensional reduction may arise from the behavior of the winding vector around exceptional points, from which one determines the winding number. For a Hermitian SSH model, there is only one point, whereas the NH approach splits it into two [41]. In the latter case, a winding vector has the chance to spin around only one of them, so it does not cover the information regarding the other exceptional point. This missed information may manifest as a dimensional reduction in the QMT, although it is not clear by which mechanism this should occur. Furthermore, this phenomenon has been recently measured experimentally in a plasmonic lattice, where the relation $g_{k_x k_x} \approx g_{k_y k_y} \approx -g_{k_x k_y}$ in the 2D momentum space has been found [29]. This completely agrees with the QMT in the *blue* region of the NH-SSH-NR model in Sec. IV (see Fig. 5 for the exact equation).

### B. Linear response to external driving

The QMT is an experimentally measurable quantity, as shown for Hermitian systems in Refs. [34,35,43]. The former work also contains a relation between the diagonal QMT components and the transition rate from an initial state to all possible final states. Inspired by that and following the methodology in Refs. [33,34,44], here we derive a similar relation for NH topological systems.

Consider a NH time-dependent system in a parameter space $\boldsymbol{\lambda}$ with the Hamiltonian given by $\hat{H}(\boldsymbol{\lambda}) = \hat{H}_0(\boldsymbol{\lambda}) + \hat{V}(\boldsymbol{\lambda})$. $\hat{H}_0(\boldsymbol{\lambda})$ denotes the unperturbed part and

$$\hat{V}(\boldsymbol{\lambda}) = \partial_{\lambda_1} \hat{H}(\boldsymbol{\lambda}^{\mathbf{0}}) \frac{2E}{\hbar \omega} \cos(\omega t) \tag{36}$$

is the periodic perturbation potential, where $E$ is the driving amplitude and $\boldsymbol{\lambda}^{\mathbf{0}}$ is the initial vector parameter at $t = 0$. If we modulate one parameter

$$\lambda_1(t) = \lambda_1^0 + \frac{2E}{\hbar \omega} \cos(\omega t)$$

with real frequency $\omega$, we find that the *integrated transition rate* reads

$$\begin{aligned}
\Gamma^{\text{int}} &= \frac{2\pi E^2}{\hbar^2} \sum_{f \neq i} \langle \partial_{\lambda_1} \psi_i^L | \psi_f^R \rangle \langle \psi_f^L | \partial_{\lambda_1} \psi_i^R \rangle \\
&= \frac{2\pi E^2}{\hbar^2} g_{\lambda_1 \lambda_1}^{LR},
\end{aligned} \tag{37}$$

where subscripts $i$ and $f$ indicate the initial and final states, respectively. Combining this formula with the dimensional reduction discussed previously results in a regime of parameters where the system does not respond to a perturbation. Next, the negative diagonal QMT components in the NH-SSH-NR model implies that the transition rate can be negative, which may be due to an inverse population. Notice that some diagonal QMT components are complex, which requires further investigations to understand whether the usual Kubo formula formalism can still be applied beyond the Hermitian and pseudo-Hermitian realm or whether a relation should be made to the real and symmetrized version of the metric. Measurements of the metric components for NH systems have thus far used real versions of the QMT [29,37].

### C. Emergent space-time, black holes, and event horizons

We noted above that pseudo-Riemannian geometries, which we have found to arise as the quantum geometries of NH systems, appear in the context of general relativity (GR) as space-time, whereas Riemannian manifolds, arising from Hermitian systems, have only space-like dimensions. Further, the quantum geometry is experimentally accessible, as the QMT appears in the form of measurable physical quantities, such as in Eq. (37). This opens up the possibility of quantum simulating space-time in the form of quantum geometries in NH systems. The study of analog gravity—that is, non-gravitational systems which exhibit fundamental properties of GR—has been a particularly fruitful area of research. This is especially true for the study of black holes, which are objects that are so heavy that outgoing light cannot escape from them beyond a certain critical radius, where the event horizon is located. Since it is impossible to perform detailed experiments on real black holes appearing in our universe, analog black holes, arising in analog gravitational systems, serve as very important testing grounds for our understanding. Black hole analogues have been found in many physical contexts, including fluid flows [45], optical systems [46], semiconductor microcavities [47], Bose-Einstein condensates [48], and superfluids [49] (see also Refs. [50,51] for a review on the topic). These experiments have provided valuable insights on topics such as black hole quasinormal modes, superradiance, Hawking radiation, black hole evaporation, and the resulting information paradox. Exploring such analog black holes seems particularly promising in the context of NH quantum geometry, as various unexpected resemblances to black holes in GR appear naturally in some of the QMT's derived in this work, which we outline below.

In the context of GR, trajectories which traverse zero metric distance, i.e., for which $ds^2 = g_{\mu\nu} d\lambda^\mu d\lambda^\nu = 0$, are the trajectories followed by massless particles and are therefore referred to as "lightlike" or "null-like." In Riemannian geometries, the metric distance is always positive semi-definite, since $g_{\mu\mu} \geqslant 0$ after diagonalizing. Conversely, pseudo-Riemannian metrics generally have negative QMT components after diagonalizing, leading to nontrivial trajectories which traverse zero metric distance. Consider the metric of the NH-SSH-NR model, for example. One can see that lightlike trajectories for these metrics correspond to lines at 45 degrees, i.e., $y = \pm z +$ constant, which is coincidentally how lightlike trajectories are typically represented in the Penrose diagrams used to study space-time [52].

The event horizon of black holes manifests as a singularity in the metric for the corresponding space-time [52], which is precisely how a phase transition manifests in the quantum metric. The fact that light cannot escape from black holes beyond the event horizon implies that the event horizon is a lightlike surface. In particular, if one considers a spherical shell of outgoing massless particles at the event horizon which are directed away from its center, this shell will remain located at the event horizon. In this sense, points on the event horizon of a black hole follow the same trajectories as massless particles, i.e., lightlike trajectories. This means that the event horizon of a black hole appears at 45 degrees in the Penrose diagram of the corresponding space-time [52]. The singularities of the QMT for the NH-SSH-NR model, where the phase transitions occur, also appear precisely at 45 degrees. By analogy with black holes, then, the phase transitions are analogous to black hole event horizon(s).[7] We thus see that NH systems broaden the scope of quantum geometry to encompass space-time, including those which exhibit the basic features of black holes mentioned above. This opens up the possibility of simulating space-time in NH systems, where concepts from GR translate to physical properties of quantum systems which may be measured in the laboratory.

## IX. CONCLUSIONS

In conclusion, we have applied a NH quantum geometry approach to different versions of the SSH model with PBC. We have successfully identified distinct topological phases and topological phase transitions in complete agreement with those found by conventional methods [4,12,41]. This confirms the validity of the quantum geometric method. Furthermore, the topological information is independent of the system size $N$, except when it is too small such that topological phenomena do not appear.

From the pseudo-Hermitian SSH model with real nonreciprocal hoppings, we first found that all QMT components are real-valued, despite the complex character introduced by the non-Hermicity. This indeed helps one to identify them with the conventional QMT. However, when we analyze the definition of eigenstates and NH QMT used [see Eqs. (5) and (17)], the resulting QMT components could be in principle complex, although the hopping parameters of the system are real. Nonetheless, this inherently depends on the definition chosen for the QMT. Currently, the authors are still unsure about which should be employed. Second, both *LL*- and *RR*-QMT components contain half of the total amount of topological information, and only *LR*-QMT components reproduce the entire picture. This is actually a generic phenomenon that happens for any NH system when considering orthonormal eigenvectors like Eq. (5), and justifies the need to use biorthogonal quantum mechanics when dealing with them [21,40]. Next, the diagonal QMT components of the *green* domain reveal that regions I and II do not belong to the same trivial phase, despite being characterized by the same two-components winding number (see Fig. 5). Last, we found that non-Hermicity gives rise to negative diagonal QMT components, which expand the workspace from the Riemannian space to pseudo-Riemannian manifolds. This opens a door to a new realm in NH topological systems.

From SSH models with complex hopping parameters, we find that the additional phase terms, compared to cases with real hopping amplitudes, directly expand the symmetric version of the QMT, $g^{\alpha\beta}_{\mu\nu}$, to a complex-valued one, where we need to use a different definition of the QMT to retrieve real components. Furthermore, since phase terms (i.e., complex number) cannot be gauged away in the quantum geometric approach, we find that they provide additional information about the system, compared to the results obtained with conventional methods by studying Hamiltonians themselves [4] (see Fig. 6). This raises the question whether some modifications of the transition rate calculations are required to study these systems, such as symmetrization; e.g., Refs. [38,53].

In general, we find that the NH approach introduces a dimensional reduction in NH systems and NH topological phases. This leads to a dark direction, where the system is unaffected by a periodic time-dependent perturbation. Recent experiments in NH systems [29] have actually observed our theoretical predictions.

In analogy to general relativity, this method also allows one to identify "event horizons" when the metric equation is zero, $ds^2 = 0$. We find that the lightlike paths coincide with the topological phase transition curves, mainly manifesting as discontinuities in the parameter space and as a divergence in the Hermitian limit.

## ACKNOWLEDGMENTS

We are grateful to E. Slootman and L. Eek for a careful reading of this manuscript, and to R. Arouca for useful discussions about the topological phase diagram of non-Hermitian SSH systems. We acknowledge the research program "Materials for the Quantum Age" (QuMat) for financial support. This program (Registration No. 024.005.006) is part of the Gravitation program financed by the Dutch Ministry of Education, Culture and Science (OCW).

## APPENDIX A: NH-SSH-NR MODEL

Figure 8 shows the 3D-graphical representation of each QMT component in Eq. (25) for $N = 200$. Specifically, their real part from the top viewpoint with the corresponding legend bar. The four phase transition lines, $z = \pm y \pm 1$, are clearly visible for all *LR*-QMT components, whereas those ones for *LL*-QMT components and *RR*-QMT components, $z = -y \pm 1$ and $z = y \pm 1$, respectively, appear as divergences according to the legend bars.

Figure 9 shows the winding of the Bloch Hamiltonian in the Pauli matrix basis, such that $H(k) = \sum_i d_i(k) \cdot \sigma_i$ with $d_x(k) = t + t_2 \cos(k)$ and $d_y(k) = t_2 \sin(k) - i\delta$ [42]. Different values of $\delta$ and $t$ are considered, with $t_2 = 1$, to produce diagrams for all topological phases together with their corresponding winding number $W = (W_1, W_2)$ [41]. The two exceptional points are given by $(\pm \text{Im}[d_y(k)], \mp \text{Im}[d_x(k)])$ in the $\text{Re}[d_x] - \text{Re}[d_y]$ representation (i.e., green and red dots).

---

[7]We mention horizon(s) because there are various black hole with multiple event horizons, including charged (Reissner-Nordström) and rotating (Kerr) black holes [52].

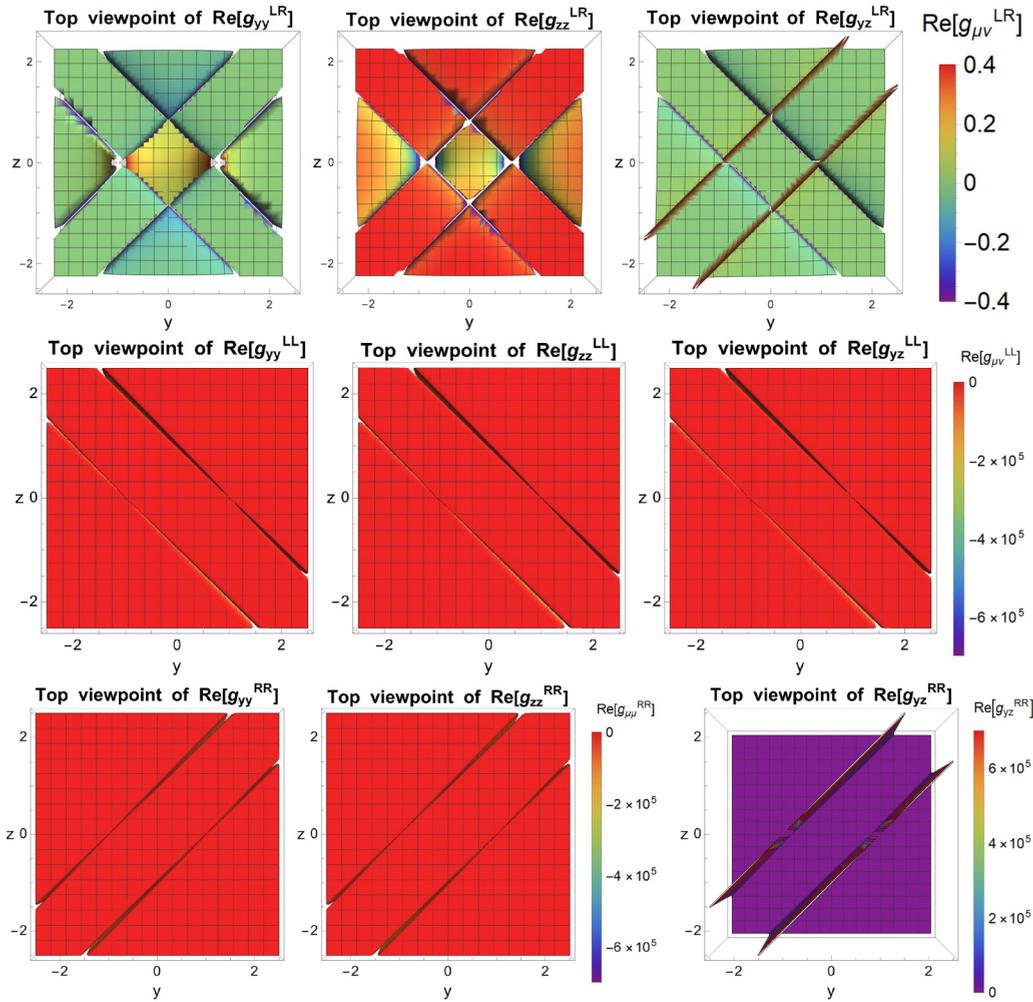

FIG. 8. $\text{Re}[g_{\mu\nu}^{\alpha\beta}]$ with $\alpha, \beta \in \{L, R\}$ in the parameter space $\lambda^\mu, \lambda^\nu \in \{y, z\} \in \mathbb{R}$ for $N = 200$. Top viewpoint.

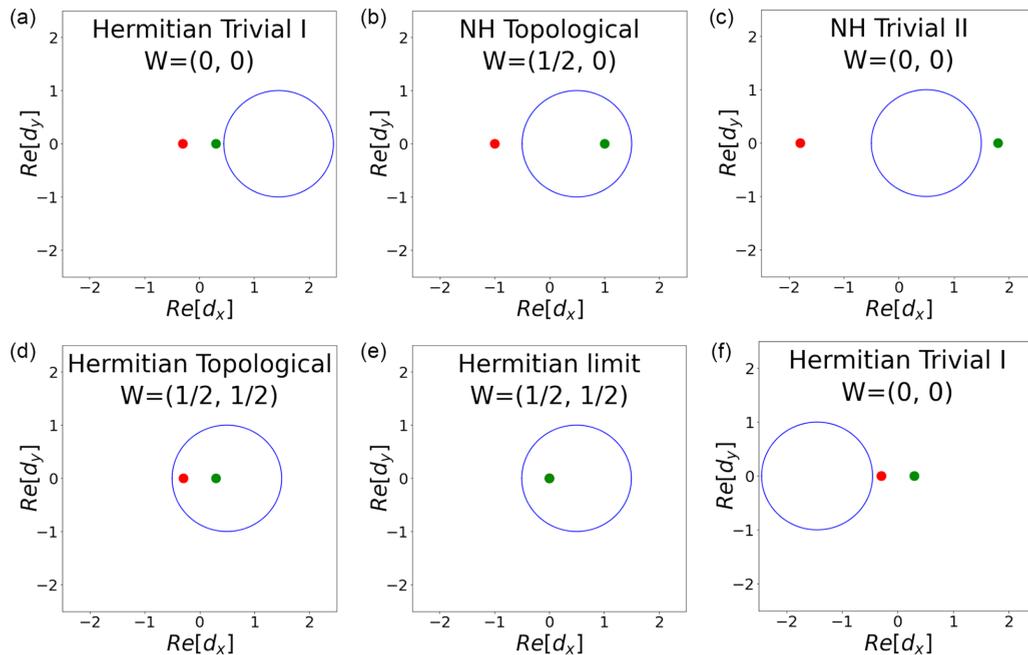

FIG. 9. The winding vector for different values of $t$, $t_2$, and $\delta$, $(t, t_2, \delta)$: (a) (1.45, 1, 0.3), (b) (0.5, 1, 1), (c) (0.5, 1, 1.8), (d) (0.5, 1, 0.3), (e) (0.5, 1, 0), (f) (−1.45, 1, 0.3). Their corresponding topological phase and winding number $W = (W_1, W_2)$ are also depicted.

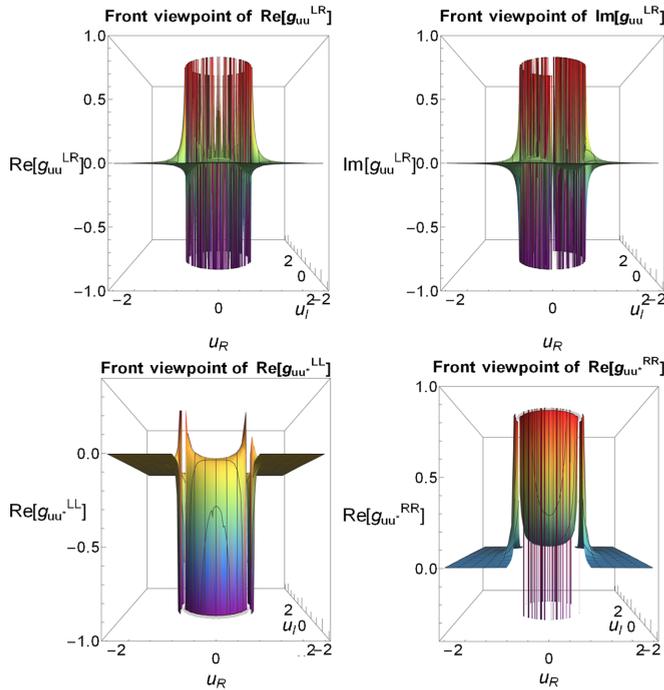

FIG. 10. Re[$g_{uu}^{LR}$], Im[$g_{uu}^{LR}$], Re[$g_{uu^*}^{LL}$], and Re[$g_{uu^*}^{RR}$] in the parameter space $u_R, u_I \in \mathbb{R}$ for $N = 200$. Im[$g_{uu^*}^{LL}$] $\approx 0$ and Im[$g_{uu^*}^{RR}$] $\approx 0$ due to finite-size effects.

### APPENDIX B: NH-SSH-C MODEL

The transformation of a QMT from the $\{u, u^*\}$-space to their real and imaginary $\{u_R, u_I\}$-space leads to Eq. (32) for *LR*-QMT components, and the *LL*- and *RR*-QMT components read

$$
\begin{aligned}
g_{u_R u_R}^{LL} &= 2\, g_{uu^*}^{LL} \in \mathbb{R}, & g_{u_R u_R}^{RR} &= 2\, g_{uu^*}^{RR} \in \mathbb{R}, \\
g_{u_I u_I}^{LL} &= 2\, g_{uu^*}^{LL} \in \mathbb{R}, & g_{u_I u_I}^{RR} &= 2\, g_{uu^*}^{RR} \in \mathbb{R}, \quad\text{(B1)} \\
g_{u_R u_I}^{LL} &= 0, & g_{u_R u_I}^{RR} &= 0.
\end{aligned}
$$

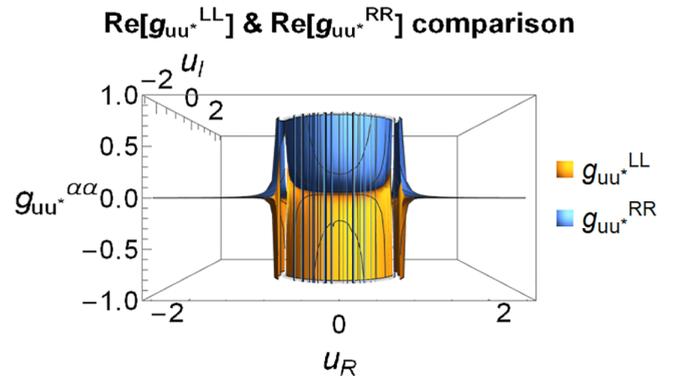

FIG. 11. Comparison between Re[$g_{uu^*}^{LL}$] and Re[$g_{uu^*}^{RR}$] in the parameter space $u_R, u_I \in \mathbb{R}$ for $N = 200$.

Figure 10 shows the 3D-graphical representation of Re[$g_{uu}^{LR}$], Im[$g_{uu}^{LR}$], Re[$g_{uu^*}^{LL}$], and Re[$g_{uu^*}^{RR}$] in the complex plane for $N = 200$, from the front viewpoint. The imaginary part of *LL*- and of *RR*-QMT components are nearly zero due to finite-size effects. It is clear that only the *LR*-QMT component is complex and both its real and imaginary parts have divergences going to $\pm\infty$. However, the *LL*-QMT component only diverges to $-\infty$, and similarly for the *RR*-QMT component, but to $+\infty$. This tells us that whereas *LR*-QMT components form a complete set by themselves, the singularity in *LL*-QMT components complements that in *RR*-QMT components, as one can observe in Fig. 11. Therefore, we conclude that *LL*- and *RR*-QMT components cover half of the total topological information each, whereas *LR*-QMT components contain the entire data. This result agrees with that found in the NH-SSH-NR model, although the topological incompleteness on *LL*- and *RR*-QMT components manifests in a distinct form, supporting the argument of the necessity to use biorthogonal quantum mechanics when dealing with NH systems.